# Statistical mechanics and Bayesian Inference addressed to the Osborne Paradox

Geoffrey Ducournau[1]

**Abstract**

Probably one of the greatest contributors of the 20th century among all academician in the field of statistical finance, M. F. M. Osborne published in 1956 [6] an essential paper and proposed to treat the question of stock market motion through the prism of both the Law of Weber-Fechner [1, 4] and the branch of physics developed by James Clerk Maxwell, Ludwig Boltzmann and Josiah Willard Gibbs [3, 5] namely the statistical mechanics. He proposed an improvement of the known research made by his predecessor Louis Jean-Baptiste Alphonse Bachelier, by not considering the arithmetic changes of stock prices as means of statistical measurement, but by drawing on the Weber-Fechner Law, to treat the changes of prices as a theoretical view of subjective perception of growth due to the information transmissible from prices' stimulus to the trader and investor. Osborne emphasized that as in statistical mechanics, the probability distribution of the steady-state of subjective change in prices is determined by the condition of maximum probability, a statement close to the Gibbs distribution conditions. Finally, Osborne argues that the best way to describe the change in prices is the same way to describe the irregular motion of particles under certain conditions. However, Osborne also admitted that the empirical observation of the probability distribution of logarithmic changes of stock prices was emphasizing obvious asymmetries and consequently could not perfectly confirm his prior theory.

The purpose of this paper is to propose an explanation to what we could call the Osborne' paradox and then address an alternative approach via Bayesian inference regarding the description of the probability distribution of changes in logarithms of prices that was thenceforth under the prism of frequentist inference. We show that the stock market returns are locally described by equilibrium statistical mechanics with conserved statistics variables, whereas globally there is yet other statistics with persistent flowing variables that can be effectively described by a superposition of several statistics on different time scales, namely, a superstatistics.

**1- Literature review:**

In one of his most fundamental paper titled "Brownian motion in the stock market" published in 1959, Osborne observed and emphasized that common stock prices and the value of money can be regarded as an ensemble of decisions in statistical equilibrium (steady-state), with properties quite analogous to an ensemble of particles in statistical mechanics under certain conditions.

One consequence of the prior statement is that the stock market price P(t + $\tau$) and P(t) are the prices of the same random choice stock at random time t + $\tau$ and t whose the traders' sensation of subjective changes in their respective price is precisely determined by the same probability distribution of a particle in a Brownian motion.

To demonstrate this consequence, Osborne will develop his argumentation through two hypotheses:
- Changes in the nominal value of prices have no significance. To be measurable and reproducible statistically, the subjective sensation of changes in prices must follow the Weber Fechner Law.
- The changes in prices of a stock are given by the conditions of trading, conditions that represent the process of decision making of the trader and investor. And this condition must verify the principle of symmetry of opportunity between buyer and seller, under which both seeks an equiprobability to maximize their profit.

[1]Geoffrey Ducournau, PhDs, Institute of Economics, University of Montpellier; G.ducournau.voisin@gmail.com

## 1.1- The subjective sensation of changes in prices and the Weber-Fechner Law

One of the fundamental assumptions of Osborne is to consider the logarithm of stock prices rather than prices itself when we are asked to investigate in a probability sense the existing relations between variables or attributes within a given population of stock prices. This assumption relies on the fact that the stimulus of price in a denominated currency, and the subjective sensation a current value of a stock price (after this stimulus) gives to the trader or investor's mind, is related to the Weber-Fechner Law. The Weber-Fechner Law stipulates that equal ratios of physical stimulus correspond to equal intervals of subjective sensations. The value of a subjective sensation, like the absolute position of an object in the physical space, is not measurable, however, changes or differences in the sensation of positions are, since by experimentation they can be reproduced and therefore fulfill the criteria of measurability. Consequently, the introduction of the Weber-Fechner law as a working hypothesis leads Osborne to examine only changes in price that occur in individual stocks (specifically the logarithms of prices ratios) since by hypothesis the absolute level of price has no significance and only changes in prices can be measured by traders or investors. Osborne gives an intuitive example [2] of the Weber-Fechner Law applied to stock prices to illustrate his rationale choice for the use of logarithms of price in preference to the price as an independent variable.

Moreover, Osborne gives another very rational reason why taking the change in logarithm prices as the level of measurement rather than prices as a nominal level of measurement by proposing a numerical example illustrating the existing symmetry of opportunity between one buyer and one seller regarding their respective process of decision making when deciding whether to buy or sell shares of a corporation. We will talk more in detail about this notion of symmetry of opportunity in 1.2.

Through this example, Osborne shows that if contrary to the hypothesis of the Weber-Fechner law, buyer and seller measure their gains by numbers of shares and numbers of dollars, then the logical decision of the buyer would be to buy, and the seller to sell, and both would legitimately expect to have a maximum probability of taking profits from their decisions due to the symmetry of opportunity, and increase the value of their initial steady state. Of course, for every single bet, both could not gain as a result of the same trial, one has to lose, one has to win. However, both could expect in the long term, after repeating independent trials of bet, to increase their initial steady state of capital. Under this condition, Osborne shows that in the long run, it is highly likely to have only winners on the market, which empirically seems to be fallacious.

However, under the Weber-Fechner Law and the hypothesis of equal interval between independent variables, gains are measured by changes in logarithms of prices. In this case, as we will show in more detail in 1.2, Osborne shows that the expected gain of each participant on the market is null since buyer and seller have the same opportunity of profit & loss and because their subjective sensation of changes in prices from the profit made by one trader and from the loss made by the other must be opposite and equal in absolute value. This is what Osborne called an indifference in the first order of decision, or logarithms changes are in a steady state of indifference. In other words, there is a statistical equilibrium between buyer and seller in the long term.

## 1.2- Analogous properties with statistical mechanics

As emphasized in the previous paragraph, Osborne's researches focus on the study and analysis of the process of decision-making of traders and investors in the stock market and arrive in the conclusion that logarithms of prices are the proper ratio to measure the sensation in changes in prices. In the same way, by analyzing theoretically and empirically the process of decision making of financial agents in the stock market, Osborne develops his main conjecture and states that once we assumed that subjective sensation of changes in prices is measured by the changes of their logarithms, it can be shown that common-stock prices and the value of money can be considered and assimilated as an ensemble of decisions in statistical equilibrium, with properties analogous to an ensemble of particles in statistical mechanics, whose the motion is determined by a similar statistical law.

---

[2] The change of a random stock price from $10 to $11 will give an objective profit of $1 when the change for another random walk from $100 to $110 will give an objective profit of $10, assessing that the difference of gain is not negligible and not equal. But, if we consider the subjective sensation of profit (or loss) as the change of logarithms prices such as $\log \$11 - \log \$10 \sim 9.53\%$ and $\log \$110 - \log \$100 \sim 9.53\%$, we assess now equal response in sensation which verify the equal interval hypothesis. Indeed, the equal interval hypothesis implies that the difference in subjective sensation of profit or loss, between $10 and $11 is equal to the subjective sensation of change from $100 to $110.

First of all, by introducing the Weber-Fechner law as a working hypothesis, it consequently leads statisticians to examine price changes of individual stock rather than the absolute level of price. This what Osborne proposes, by statistically studying as example the histograms of the distribution function of $\log(P)$ for common stocks that belong to the Exchange NYSE during the year 1956. Osborne found that the accumulated distributions for intervals of a month and a year are both nearly normal in ratio units for the NYSE Exchange. This first assessment of normality in the changes of the logarithm of prices suggest that it may be a consequence of many independent variables contributing to the changes in values (as defined by the Weber-Fechner Law), and Osborne states that the normal distribution also arises in any stochastic processes involving the large number of independent variables as the probability distribution for a particle in a Brownian motion.

Secondly, Osborne proposes to define a logical decision-making process of traders and investors when deciding whether to buy or sell common stock in order to theorize the previous empirical assessment regarding the normality of changes of logarithms' prices.

For this purpose, Osborne assumes first a trader willing to increase his current capital by buying a certain number of shares at time t for a given price $P_0(t)$. This trader is consequently dealing with two possibilities, whether buying for future sell at $t + \tau$ (choice A) or not buying (choice B). Since according to the Weber-Fechner law the quantity measurable by the traders 'mind is the subjective sensation of price changes, Osborne defines by $Y_A(\tau) = \Delta \log(P(t)) = \log \frac{P(+\tau)}{P_0(t)}$ the possible change in the logarithms of the price for choice A and $Y_B(\tau) = 0$ corresponding to the null change with probability one regarding the choice B. Consequently, Osborne understands that the decision-making of the trader to buy or not is determined by whether the estimated expected value of $Y_A(\tau)$ is positive or negative. And Osborne defines this expectation as $\rho(Y_A)$, namely the probability of possible outcome $Y_A(\tau)$.

Reciprocally, to make this decision possible, a seller must at the same time as the buyer develops the exact opposite process of decision making. Therefore, Osborne supposes that for the buyer, his estimate of the expectation value for the change in value $\Delta \log P$ for the stock is positive while the seller's estimate for the same value of change is negative.

Consequently, accordingly, to this process of decision making, we have an equality of opportunity on bidding between buyers and sellers, and it appears that the most probable condition under which a transaction become possible and recorded, is obtained when these two estimated from the buyer and the seller are equivalent and opposite namely: $E(\Delta \log P)_{Buyer} + E(\Delta \log P)_{Seller} = 0$, where P is the price per share, and $E$ is the estimate of the expectation value of the price. Even though decision-making process can be much more complicated than this example, and may involve several alternatives and sequences of decisions in which the estimate of the probabilities may depend on multivariate problems, the general approach remains the same, namely based on given and available information, the purpose is to maximize estimates of probabilities of the best expectation value of the end result.

Hence, Osborne concludes that the condition under which the estimated change in price of the market as a whole (consisting of buyers and sellers) is the most probably to be recorded is $E(\Delta \log P)_{M=B+S} = 0$.

In other words, Osborne states that the contestants are unlikely to trade unless there is equality of opportunity to profit.

Thus, in regard to this prior conclusion, Osborne question himself about the effect of this condition of equality of opportunity to profit on the distribution function of $\Delta \log P$. Osborne argues that the answer given by Gibbs regarded the distribution of an ensemble of molecules in equilibrium can be applied to the distribution of change in prices of stock. Indeed, Osborne postulates that the actual distribution function is determined by the condition of maximum probability.

Moreover, this condition remains true when traders or investors must decide between two stocks, whether buying stock A or stock B. The rational decision process relies on estimating (in any sense) the possible outcomes of both stocks namely, $\forall i \in [0, 1, ..., n], Y_A = \sum_i Y_{Ai}$ and $Y_B = \sum_i Y_{Bi}$ with respective probabilities $\rho(Y_A) = \sum_i \rho(Y_A)_i$ and $\rho(Y_B) = \sum_i \rho(Y_B)_i$. Then the logical choice is to choose the stock for which the expectation value $E$ of the outcome namely $E(A) = \sum_i Y_{Ai} * \rho(Y_A)_i$, $E(Y_B) = \sum_i Y_{Bi} * \rho(Y_B)_i$ is the larger, namely the stocks that maximize the probability of profit.

Osborne then generalizes the previous statement by assuming that the decisions for each transaction in the sequence of transactions in a single stock are made independently (in the probability sense), and by considering the condition of maximum probability stipulated above, he expects that the distribution function for $Y_\tau = \log \frac{P(+\tau)}{P_0(t)}$ is normal with zero mean and a dispersion $\sigma_{Y_\tau}$ which increases as the square root of the number of transactions (transactions fairly distributed in time) namely will increase as the square root of the time interval such as $\sigma_{Y_\tau} = \sigma\sqrt{\tau}$, where $\sigma$ is the dispersion at the end of unit time. Therefore, for a large number of random transactions in the sequence, the central limit theorem assures that $Y_\tau$ will approach a normal distribution function whatever the underlying distribution of the stock price.

Thus, this last consideration has a strong analog in the diffusion of a molecule undergoing collisions with its neighbors. Indeed, whatever the physical law (deterministic or not) at the origin of the motion of the molecules, the distribution of the position of a particle initially (at time t) locate at some point in space and time, will also increase as the square root of the time interval $\tau$ after t.

### 1.3- *Osborne paradox on the description of the stock return probability distribution.*

Based on the main prior assumptions that we have emphasized previously, namely the fact that Osborne considers as a necessary condition to have an equality of opportunity in bidding between buyers and sellers to be able to record a transaction, he consequently shows that the long term means expectation of price is null. Then, Osborne concludes that the impact of this prior assumption on the distribution function of $\Delta \log P$ is determined by the maximum probability, will be normal with zero mean and with a dispersion $\sigma_{Y(\tau)} = \sigma\sqrt{\tau}$, which increases as the square root of the number of transactions.

However, empirically speaking, Osborne also assesses a secular increase of the expectation stock index NYSE prices with increasing time interval $\tau$, at a certain rate per year, increasing the fluctuation of the price $P$, and being at the origin of obvious asymmetries within the change in log-returns of prices. Assessment that contradicts partially his deduction from his prior assumption which consists of having long-term expectation mean of changes in price equal to zero. And Osborne adds that this increase has not as causes the long-term economical inflation that could be at the origin of assets inflation since the expected number of shares purchasable in the future should also increase with time interval $\tau$ in an identical way that the changes in prices.

So how to address this paradox? This is what we will try to answer within the next part.

### 2. Bayesian thinking addressed to the Osborne Paradox

As we have seen previously, the Osborne paradox relies on the fact that his conclusion on the steady-state distribution function of the change in logarithms of prices as being determined by the maximum probability is not corroborated by the empirical observation.

Our assumption on the origin of this paradox comes from one assumption made by Osborne regarding the equality of opportunity of profit between both buyers and sellers. It is indeed rational to assume that a trader or investor will decide to record a transaction only if his subjective sensation of maximizing his probability of success is fulfilled. However, is it reasonable to consider that this opportunity of profits as being equals between agents? Since it is obvious that traders are heterogeneous, is it not more conceivable to consider their respective maximum probability of success as being also heterogenous?

## 2.1- Thought experiment: The Osborne's Demon

In this part, we propose to make a thought experiment example to illustrate the Osborne paradox namely the asymmetry of opportunities between buyers and sellers.

However, just before we wish to define some important concepts and definitions related to statistical mechanics and thermodynamics.

*The second law of Thermodynamics*

One of the most fundamental and probably the most universal regulators of natural activity known in science is owed by the second law of Thermodynamics. The second law of thermodynamics states that, in a closed system, the entropy does not decrease. That is, if the system is initially in a low entropy state, its condition will tend to slide spontaneously toward a state of maximum entropy (complex system). For example, if two blocks of metal at different temperatures are brought into thermal contact, the unbalanced temperature distribution rapidly decays to a state of uniform temperature as energy flows from the hotter to the colder block. Having achieved this state, the system is in equilibrium. The second law also tells that the entropic progress from simple system to complex systems may then be viewed as an example of the general tendency of chaotic disruptions to disturb organization and structure. Any system that is subject to random agitations will eventually attain its most chaotic condition.

*Entropy*

The entropy measures the irreversibility of change. When the entropy reached its maximum, then the system is in perfect equilibrium and will remain unchanged over time. Change in entropy can only increase or equal to zero when the system has reached its equilibrium.

An isolated and conservative system with no interaction with the surrounding will fatally tend to its maximum entropy and equilibrium.

Only dissipative systems that interacts with its surrounding can change and reduce the value of entropy of the system. But to do so we have to increase the entropy of the surrounding. This important concept because any open and dissipative system will be in non-equilibrium since they will never perfectly reach their maximum entropy.

In statistical mechanics the entropy $S$ is given by the Boltzmann's equation as follows:
$S = k_B \ln(\omega)$ where $k_B$ is the Boltzmann constant and $\omega$ the number of combinations of microstates corresponding to the macro-state $\Omega$.

Let's now imagine an Osborne's demon able to know every micro-information about traders and investors, such as their probability of profit, their frequency of trade, their time frame investment, and so on…

**Thought experiment number 1: Conservative system**

Knowing the presence of Osborne's demon, let's now define the following context. Let's imagine that we are in presence of an ensemble of participants, an ensemble that we will call a macro-state $\Omega$, and we consider initially each participant homogeneous, having the same probability of opportunity. We describe each participant as being a micro-state $\mathcal{M}_i$ of $\Omega$, with a respective probability $p_{\mathcal{M}_i}$, and $i = 1, \ldots, n$, with $n$ the number of micro-states.

Because the Osborne's demon knows the micro details of each $\mathcal{M}_i$, he knows that their process of decision making is not frequentist by keeping the identical likelihood of profit after every bet, but rather Bayesian. Consequently, after observing the outcome from their bet, the Osborne's demon will update his belief regarding the prior likelihood of each $\mathcal{M}_i$ based on the outcome of their previous bet. Thus, maximizing the opportunity of each $\mathcal{M}_i$ rely on maximizing their posterior probability of observing a profit such as:

$$P(\text{win}|\mathcal{M}_i) = \frac{P(\text{win}) * P(\mathcal{M}_i|\text{win})}{P(\text{win}) * P(\mathcal{M}_i|\text{win}) + P(\text{loss}) * P(\mathcal{M}_i|\text{loss})}$$

Now that the context is defined, we propose the following illustration of the first thought experiment. Figure 1 emphasizes six successive macro-states in which a transaction between at least two micro-states occurs. Each macro-state is constituting of five independent micro-states {A, B, C, D, E} and we assume in step 1 that we are under the Osborne condition, namely each participant have the same likelihood of profit and is maximum.

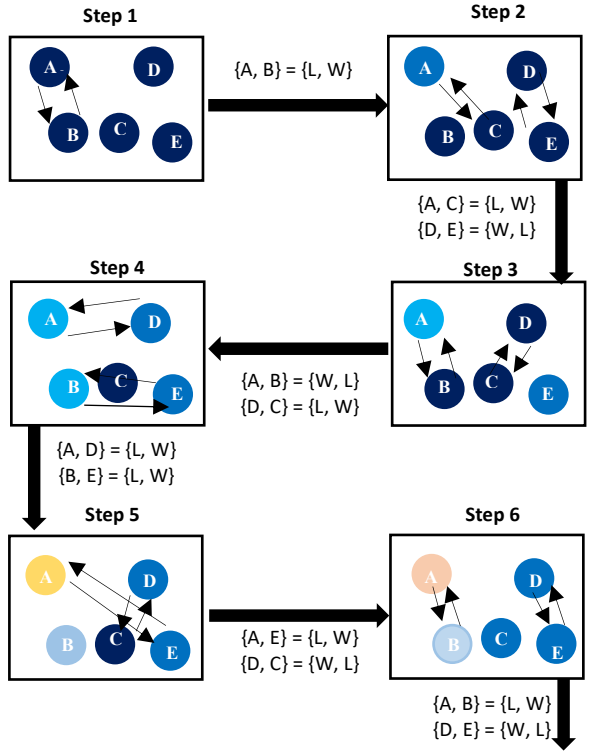

**Figure 1:** Changes in the likelihood of profit of micro-states due to update of credence in prior belief based on bet's outcomes, leading to changes in the posterior probability of micro-states.

Figure 1 must be read by following the array, the transaction that occurs in step 1 is following by the transaction occurring in step 2, and so on…

We represent the posterior probability by a given color, if the color is the same for different micro-states, it means they are homogeneous.

We can assess that the color of micro-states is changing as the step progress. The reason is that contrary to the Osborne condition that adopts a frequentist inference, we assume here that each micro-state is Bayesian and the Osborne's demon updates the posterior probability of each micro-state step after step. The following Table1 gives a picture of the macrostate's situation at step1. The posterior probability $P(\text{win}|A) = P(\text{win}|B) = P(\text{win}|C) = P(\text{win}|D) = P(\text{win}|E) = 1$, and therefore we respect the Osborne condition of equiprobability of opportunity.

| Step 1 | A | B | C | D | E | |
|---|---|---|---|---|---|---|
| Win | 1 | 1 | 1 | 1 | 1 | 5 |
| loss | 0 | 0 | 0 | 0 | 0 | 0 |
| $P(\text{win}|\mathcal{M}_i)$ | 1 | 1 | 1 | 1 | 1 | 1 |

**Table 1:** Representation of the number of wins and losses of each microstate at step 1.

However, since the marginal probability $P(\text{win}) = 0.5$, for each bet, one micro-state must lose and the other must win. In figure 1 we assess that after the bet occurred in step1, the Osborne's demon knows that A lost against B. Consequently, since A is a Bayesian, he must update his likelihood of observing itself as a winner for the next bet. This is assessable within table 2 where the posterior probability of profit of {A} is updated to 0.9. As for {B}, its posterior probability remains the same even though its likelihood of being the next winner increases.

| Step 2 | A | B | C | D | E | |
|---|---|---|---|---|---|---|
| Win | 1 | 2 | 1 | 1 | 1 | 6 |
| loss | 1 | 0 | 0 | 0 | 0 | 1 |
| $P(\text{win}|\mathcal{M}_i)$ | 0.5 | 1 | 1 | 1 | 1 | 0.9 |

**Table 2:** Representation of the number of wins and losses of each micro-state at step 2.

In step 2 Figure shows that the couples {A, C} and {D, E}, bets against each other. {A} and {E} are both loosing. Table 3 emphasized the respective update of their posterior probability.

| Step 3 | A | B | C | D | E | |
|---|---|---|---|---|---|---|
| Win | 1 | 2 | 2 | 2 | 1 | 8 |
| loss | 2 | 0 | 0 | 0 | 1 | 3 |
| $P(\text{win}|\mathcal{M}_i)$ | 0.16 | 1 | 1 | 1 | 0.27 | 0.69 |

**Table 3:** Representation of the number of wins and losses of each microstate at step 3.

The remaining changes in posterior probability as step progress are emphasized within Table4.

| Step 4 | A | B | C | D | E | |
|---|---|---|---|---|---|---|
| Win | 2 | 2 | 3 | 2 | 1 | 10 |
| loss | 2 | 1 | 0 | 1 | 1 | 5 |
| $P(\text{win}|\mathcal{M}_i)$ | 0.33 | 0.5 | 1 | 0.5 | 0.33 | 0.53 |
| Step 5 | A | B | C | D | E | |
| Win | 2 | 2 | 3 | 3 | 2 | 12 |
| loss | 3 | 2 | 0 | 1 | 1 | 7 |
| $P(\text{win}|\mathcal{M}_i)$ | 0.28 | 0.37 | 1 | 0.6 | 0.54 | 0.56 |
| Step 6 | A | B | C | D | E | |
| Win | 2 | 2 | 3 | 4 | 3 | 14 |
| loss | 4 | 2 | 1 | 1 | 1 | 9 |
| $P(\text{win}|\mathcal{M}_i)$ | 0.24 | 0.39 | 0.66 | 0.72 | 0.66 | 0.53 |
| Step 7 | A | B | C | D | E | |
| Win | 2 | 3 | 3 | 5 | 3 | 16 |
| loss | 5 | 2 | 1 | 1 | 2 | 11 |
| $P(\text{win}|\mathcal{M}_i)$ | 0.22 | 0.51 | 0.67 | 0.77 | 0.51 | 0.54 |

**Table 4:** Representation of the number of wins and losses of each microstate at step 4/5/6/7.

Thus, we understand by updating the posterior probability of profit of micro-states that the Osborne's demon changes the future expected macro-sate gain since we do not assess anymore equiprobability of opportunity for each bet. Consequently, locally and temporally, it is assessable that the expected mean of stock returns (sum of two opposite microstates posterior probability of profit) will differ from zero because of the disparities of the probability of profit between agents. And the number of combinations where the stock returns will differ from zero is equal to the number of combinations of pairs or micro-states with a different posterior probability of profit.

We can formally say that a $k = 2$ pair of a set $\Omega$ of elements $\mathcal{M}_i$ with $N$ the number of $\mathcal{M}_i$ is equal to the binomial coefficient $\binom{N}{k} + \sum \binom{N-k}{k} = \frac{N!}{k!(N-k)!} + \frac{(N-k)!}{k!(N-2k)!}$, if and only if $N - k > 2$, else it is only equal to $\binom{N}{k}$.

For example, in step 5 we have three heterogenous and distinguishable micro-states ({A}, {B}, {C}) and two homogeneous micro-states {D} and {E}. By assuming that the order of selection does not matter since micro-states are independent, we can deduct that the number of potential expected returns resulting from step 5 is equal to $\binom{5}{2} + \binom{3}{2} = 10 + 3 = 13$.

However, this assessment does not reject Osborne's hypothesis that in the long term, the expected mean of stock market returns tends to zero. Indeed, we only show that by assuming inequality of opportunity between agents, we can assess dispersion on stock market returns. This is yet a good statement because it can partially explain the Osborne Paradox regarding the existence of the stock market returns' asymmetries but cannot explain however why we assess strong asymmetries.

We propose to numerically compute this first thought experiment to emphasize the previous assessment with an initial number of microstates equal to 50 (see Figure 2).

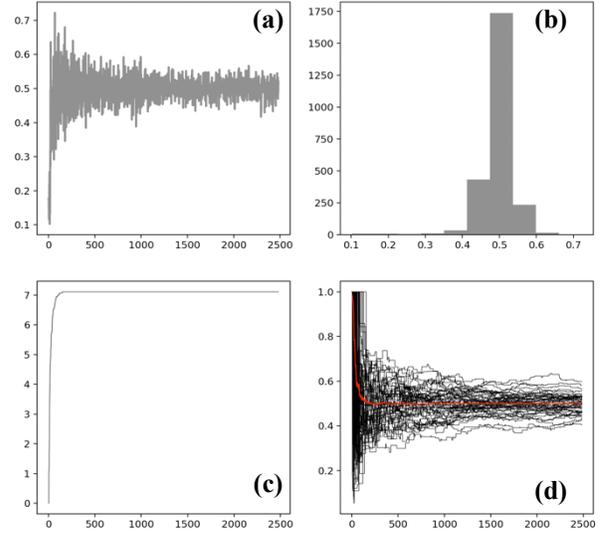

**Figure 2:** (a) Macro-state smoothing Posterior probability, (b) Macro-state Posterior probability distribution, (c) Entropy, (d) Microstates respective posterior probability (black line) and Mean of posterior probabilities of the microstate's ensemble (red line).

As emphasized by Figure2, at the beginning of the experimentation, every single microstate starts under the Osborne condition and has an equal opportunity of profit. Their likelihood and posterior probability equal 1 since when they start to bet on the market, they technically have never lost a trade yet. As interactions between microstates progress, we can assess that respective posterior probability become quickly disperses (Figure2, (a), (d)) and that the entropy of the Macro-state is increasing due to the number of increasing combinations of pair of microstates available (Figure (c)) within the phase space. However, even though as we can see in Figure2, (d) each microstate has inequal and disperses posterior probabilities, we also assess that in the long term, the expected mean of the posterior probability of the ensemble of microstates, tend to a certain equilibrium, tend to a mean posterior probability of 0.5. Thus, in the long term, the macro-state posterior probability converges to the Osborne marginal probability of 0.5. And this value has been reached when the macro-state's entropy has reached its maximum.

Consequently, even under the assumption that traders and investors have unequal maximum opportunity of profit due to Bayesian inference, computer simulations show that the Osborne theory remains valid. In the long term, the respective unequal posterior probability of profit of each trader and investor tends to an equilibrium which is the natural marginal probability of profit and loss, namely 0.5.

The stock market returns follow the second law of Thermodynamics.

And finally, the dispersion within the stock market returns is explained by local and temporal inequality of profit between microstates, until the ensemble of microstates has maximized its entropy and reach its equilibrium.

However, under the previous thought experiment and Osborne conditions, we are still making a strong assumption; that the stock market returns are a conservative and isolated system with no interactions with its surrounding. By saying isolated, we mean that over time, we assess no new injection of surrounding microstates interacting with current and previous microstates, and we assess no subtraction of current microstates that decide to leave the market due to a probability of profit too weak. And it is therefore obvious that the system will reach its equilibrium due to the second law of Thermodynamics.

The next question that we wish to answer is, how the macro-state's long term mean of probability of profit & loss evolve when we consider that over time the system is dissipative with constant interaction with its surrounding, with constant exchange and interaction of new microstates willing to join or exit the market?

To answer this question, we propose a second thought experimentation in which the stock market returns as macro-state is not considering as a conservative system anymore but considering as dissipative.

**Thought experiment number 2: Dissipative System**

Let's now consider another thought experiment by considering now the macro-state constituting of an inhomogeneous ensemble of microstates. In other words, we will consider local macro-states $(\omega_1, \omega_2, \omega_3 \ldots \omega_N)$ more or less heterogeneous with each other, constituting of respective microstates of different size such as $\mathcal{M}_1 \subset \omega_1$ and $\mathcal{M}_1 = \{\mathcal{M}_{1i}\}$ with $i = 1, \ldots, n_1$   $n_1$ corresponds to the number of microstates $\mathcal{M}_1$. The ensemble of local macro-states $\omega_i$ constitutes the stock market returns macro-state $\Omega$.

Thus, we consider the stock market returns as a macro-state that exhibit spatio-temporally inhomogeneous dynamics, due to the fact it is constituting of an ensemble of local macro-states that represent an ensemble of coarse-grained.

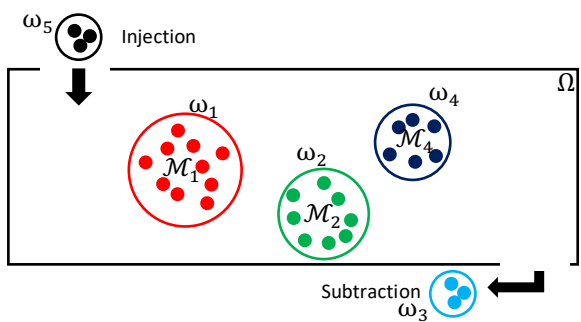

**Figure 3:** Dissipative macro-states system with coarse-gains constituents.

As represented in Figure 3, each coarse-grained have a different size and color, because we are willing to emphasize their heterogeneities. Indeed, it represents individual or group of traders, investors that share similar investment strategies, investment scale, and so on, and consequently, when microstates of these respective coarse-grained interacts with each other, they are under the same condition that emphasized within the previous thought experiment. They are more or less an isolated system that tends to equilibrium and maximized their entropy. They, therefore, have a similar infinite posterior probability of opportunity, but the fluctuation of this posterior probability differs from each other since their size, and the type of their microstates are different.

Consequently, we understand that we are in presence of inhomogeneous local macro-sate that all tend to equilibrium because of the Osborne principle (more generally because of the second law of Thermodynamics), however, because the system is dissipative with constant injection and subtraction of new local macro-states, it tends to never reach an equilibrium in terms of coarse-grained homogeneity.

How this assessment impact the overall macro-state $\Omega$ ?

To answer this question, we propose to also simulate this thought experiment by computer. We propose to simulate four different local macro-states $\omega$ respectively constituting of 750, 225, 150, and 425 micro-states. Because the model cannot run infinitely for an infinite number of local macro-states, the global macro-sate $\Omega$ will tends to equilibrium. But the principle here is to understand how the ensemble evolves as a whole statistically speaking.

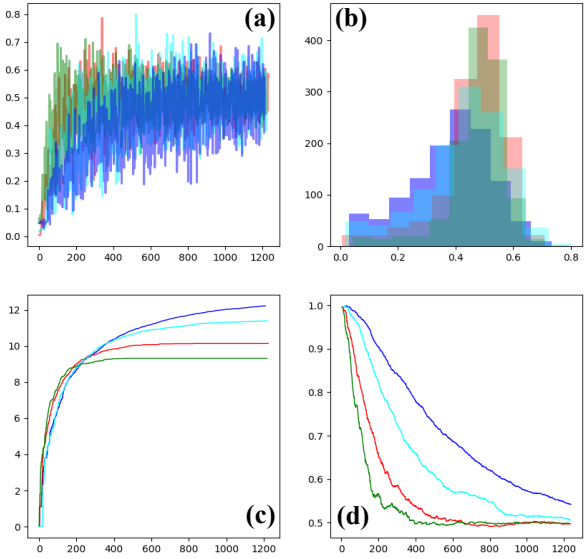

$\omega_1 : \mathcal{M}_1 = 750$ / $\omega_2 : \mathcal{M}_2 = 225$ /
$\omega_3 : \mathcal{M}_3 = 150$ / $\omega_1 : \mathcal{M}_4 = 425$

**Figure 4:** (a) Coarse-grained smoothing posterior probability, (b) Coarse-grained posterior probability distribution, (c) Coarse-grained entropy, (d) Respective coarse-grained mean of posterior probabilities of the microstate's ensemble.

Figure4 emphasized the simulation outcomes of the second thought experiment. The first thing we can assess is that each coarse-grained sooner or later tend to their equilibrium namely a posterior probability of 0.5. The speed of convergence depends on the number of microstates, the larger it is, the longer it takes to converge (Figure2 (d)). Moreover, as previously we assess that the equilibrium is reached when their respective entropy is maximized.

However, an interesting thing exhibit in Figure2 (a) and (b). Indeed, we can assess that each Coarse-grained presents a certain dispersion of posterior probability once again due to their Bayesian inference. But this dispersion is not the same from one to another. It seems that if each coarse-grained tend to the same infinite equilibrium, the variance of their posterior probability is different. Each $\omega_i$ have their own and respective $\sigma_i^2$ but the same long term mean equilibrium of opportunity (0.5).

This very important assessment is even more relevant when we analyze its consequence on the macro-state $\Omega$ probability distribution function. Figure2 (b) gives us a good understanding of what happening. We have a superposition of inhomogeneous probability distribution function of local macro-state that all tend to center to the same mean but at different scales, and with different variance. By superposing this ensemble of probabilities distribution function, we obtain a non-Gaussian distribution with strong asymmetries and fat tails.

Because each coarse-grained are different in space and time to another, and that each Coarse-grained are more or less far from equilibrium, they do not describe exactly the same centered distribution with the same statistics metrics (some are closer to the final posterior probability equilibrium) but all tend to reach equilibrium. And by superposing this ensemble of statistics, we obtain a not symmetric distribution that tends to be centered to the marginal probability 0.5, as Osborne deducted half-century ago.

And this convergence to the marginal probability depends on how the macro-state being studied is dissipative.

**Summary of the Osborne Paradox:**

The stock market returns could be considered as a macro-state that exhibits spatio-temporally inhomogeneous dynamics. Indeed, it is constituting of an ensemble of local macro-states called Coarse-grained each in local equilibrium (or tend to local equilibrium) but with different variance of the posterior probability.

We show that the cause of this inhomogeneity comes from the fact the stock market returns is a dissipative system (more or less far from equilibrium) with constant rotation and update of constituents by new injection and subtraction of Coarse-grained with respective and different statistics metrics.

Finally, we understand that locally, the stock market returns are described by equilibrium statistical mechanics with conserved statistics variables that tend to equilibrium, whereas globally there is yet other statistics with persistent flowing variables that can be effectively described by a superposition of several statistics on different time scales, namely, a superstatistics [7, 8]. This approach involves the existence of intensive parameters Θ that fluctuate on a much larger time scale than typical relaxation [3] time of the local dynamics. In Finance, it can be interpreted as an effective friction constant due to inhomogeneous dynamics such as fluctuating volatility.

---
[3] Return of a perturbed system into equilibrium.

Finally, the answer to the Osborne paradox is that due to the non-conservative system, the dispersion of the stock market returns is not defined as Osborne stated by $\sigma\sqrt{\tau}$ with constant $\sigma$, but by $\sigma_i\sqrt{\tau}$ with $\sigma_i$ the stock volatility being a random variable described by a given distribution with conserve parameter Θ.

The superposition of statistics that gives Θ, is not additive but multiplicative (geometric system) and tends more or less fast to a posterior probability equilibrium as steady state (ergodic system). This superposition of statistics exhibits non-Gaussian behavior with fat tails, which can decay with a power-law or in a more complex way [2].

We thus propose to give an update to the Osborne definition by stating that:

*"The stock market returns can be considered as a geometric ergodic non-equilibrium system that tends to converge to a steady state and maximized its entropy, without completely reaching it".*

The Osborne paradox relies on the fact that he considered the stock market returns as a conservative system, whereases it is a dissipative one.

In the next sub-parts of this article, we will propose a Bayesian Inference to estimate the parameter Θ.

### 2.2- A Bayesian alternative to the Osborne Paradox

If we consider the Osborne conclusion and take into account his probability distribution function of the stock market returns based only on the maximization of probability, we understand that Osborne is actually only maximizing the likelihood of the distribution of $\Delta \log P$ knowing that the parameter Θ of the distribution is determined by a given model which is a Normal distribution of mean zero and variance $\sigma^2$, namely maximizing the probability of observing the data $\mathcal{D}_n = X_1, \dots, X_n$ of n observations with $\Delta \log P_i = X_i$ conditional on a deterministic parameter $\theta$ and a model $\mathcal{M} = N(0, \Theta)$; in short maximizing $p_\Theta(\mathcal{D}_n|\mathcal{M})$ where $\theta$ is viewed as a deterministic value.

However, since this likelihood does not take into account the potential asymmetry within the distribution due to the traders and investors heterogeneities, consequence as we previously showed, of a lack of information regarding the stock market variance which is considering as constant in Osborne analysis due to the fact that Osborne follows a Frequentist approach. Therefore, it seems interesting to propose another inference regarding the understating that we can have on the distribution and propose consequently a prior belief on the distribution that could best describe the stock volatility distribution and interpret this subjective belief in terms of Bayesian probabilities in order to propose a posterior distribution that could the best describes the stock market returns distribution.

We propose the following alternative Bayesian inference procedure:
- We choose a probability density $\pi(\Theta)$ as the prior distribution that expresses our beliefs about a parameter Θ before we observe any data.
- We then choose a statistical model $p(x|\Theta)$ that reflects our beliefs about x given Θ.
- Finally, after observing the data $\mathcal{D}_n = \{X_1, \dots, X_n\}$, we update our beliefs and calculate the posterior distribution $p(\Theta|\mathcal{D}_n)$.

By Bayes 'theorem, the posterior distribution is given by:

$$p(\Theta|\mathcal{D}_n) = \frac{\pi(\Theta) \cdot p(\mathcal{D}_n|\Theta)}{p(\mathcal{D}_n)} = \frac{\mathcal{L}_n(\Theta)\pi(\Theta)}{c_n} \propto \mathcal{L}_n(\Theta)\pi(\Theta) \quad (2)$$

Where $\mathcal{L}_n(\Theta) = \prod_{i=1}^{n} p(X_i|\theta)$ is the likelihood function and:

$$c_n = p(X_1, \dots, X_n) = \int p(X_1, \dots, X_n|\theta)\,\pi(\theta)d\theta.$$
$$= \int \mathcal{L}_n(\Theta)\pi(\Theta)d\Theta \quad (3)$$

is the normalizing constant, which is also called the evidence.

Thus, if we use the same prior belief of Osborne regarding the fact that the probability of observing the data $\mathcal{D}_n$ is determined by a Gaussian model, in our case we will consider a known mean $\mu$ and consider the free parameter as being the variance $\sigma^2$, we can consequently write the likelihood function given by the Bayes' theorem as:

$$p(X_1, \dots, X_n|\sigma^2) \propto (\sigma^2)^{-\frac{n}{2}} \cdot \exp\left(\frac{-1}{2\sigma^2}\sum_{i=1}^{n}(X_i - \mu)^2\right)$$
$$= (\sigma^2)^{-\frac{n}{2}} \cdot \exp\left(\frac{-1}{2\sigma^2}n\overline{(X - \mu)^2}\right) \quad (4)$$

Where $(X - \mu)^2 = \frac{1}{n}\sum_{i=1}^{n}(X_i - \mu)^2$

From equation (4) we assess that the probability distribution function of the likelihood belongs to the exponential family distributions, where the sample $\mathcal{X} = \mathbb{R}_+$ is the non-negative real line and we can simplify the above likelihood distribution by:

$p(x|\Theta) = \Theta e^{-x\Theta}$ (5).

The conjugate prior $\pi(\Theta)$ of this probability likelihood, in the most convenient parametrization, is the inverse Gamma distribution given by:

$\pi_{\alpha,\beta}(\Theta) \propto \Theta^{-(\alpha+1)} e^{-\beta/\Theta}$.

With this prior, the posterior distribution of $\sigma^2$ is given by:

$\sigma^2|X_1,\ldots,X_n \sim \text{InvGamma}\left(\alpha + \frac{n}{2}, \beta + \frac{n}{2}\overline{(X-\mu)^2}\right)$

and is also an inverse gamma distribution function. Therefore, we can assess the difference between the Osborne and Bayesian inference. Rather than only having a belief on the probability distribution of the likelihood of observing data given a model and its known parameters, we conjugate this likelihood belief with a prior belief on the probability distribution of the model's parameters.

This Bayesian approach have been recently confirmed in the analysis of the probability distribution of volatility and stock returns on the SP500. Indeed, Tao Ma and R.A [9]. Serota works have demonstrated that the Generalized Inverse Gamma (GIGa($\alpha, \beta, \gamma$)) provides the best fit to VIX/VXO volatility distribution function and that the product GIGa($\alpha, \beta, \gamma$) * $N(0,1)$ provides the best fit to the stock returns distribution.

We will propose in future research, a Bayesian inference approaches to measure how strong is the credence we have regarding the prior belief we stated on the probability distribution model of the parameter $\Theta$ as on the choice of the model of the likelihood probability distribution.

To this end, we will consider $K$ parametric models $\mathcal{M}_1,\ldots,\mathcal{M}_K$ and the observable data $\mathcal{D}_n = \{X_1,\ldots,X_n\}$. We then assign a prior probability $\pi_j = \mathbb{P}(\mathcal{M}_j)$ to model $\mathcal{M}_j$ and prior $p_j(\Theta_j|\mathcal{M}_j)$ to the parameters $\theta_j$ conditioned to the model $\mathcal{M}_j$. Given the prior probability $p_j(\Theta|\mathcal{M}_j)$, the posterior probability of model $\mathcal{M}_j$ conditional on data $\mathcal{D}_n$ is:

$$p(\mathcal{M}_j|\mathcal{D}_n) = \frac{p(\mathcal{D}_n|\mathcal{M}_j).\pi_j}{p(\mathcal{D}_n)}$$
$$= \frac{p(\mathcal{D}_n|\mathcal{M}_j).\pi_j}{p(\mathcal{D}_n|\mathcal{M}_j).\pi_j + p(\mathcal{D}_n|\overline{\mathcal{M}_j}).\overline{\pi}_j} \quad (6)$$

With:

$$p(\mathcal{D}_n|\mathcal{M}_j).\pi_j + p(\mathcal{D}_n|\overline{\mathcal{M}_j}).\overline{\pi}_j = \sum_{k=1}^{K} p(\mathcal{D}_n|\mathcal{M}_k).\pi_k$$

Hence:

$$p(\mathcal{M}_j|\mathcal{D}_n) = \frac{p(\mathcal{D}_n|\mathcal{M}_j).\pi_j}{\sum_{k=1}^{K} p(\mathcal{D}_n|\mathcal{M}_k).\pi_k} \quad (7)$$

Where: $p(\mathcal{D}_n|\mathcal{M}_j) = \int \mathcal{L}_j(\Theta_j) p_j(\Theta_j) d\Theta_j$, and $\mathcal{L}_j$ is the likelihood function for the model $j$.

Consequently, we can write the following equality:

$$\frac{\mathbb{P}(\mathcal{M}_j|\mathcal{D}_n)}{\mathbb{P}(\mathcal{M}_k|\mathcal{D}_n)} = \frac{\mathbb{P}(\mathcal{M}_j|\mathcal{D}_n)\pi_j}{\mathbb{P}(\mathcal{M}_k|\mathcal{D}_n)\pi_k} \quad (8)$$

Thus, since our purpose is to choose the best fit model between $\mathcal{M}_j$ and $\mathcal{M}_k$ with $k \in K$, the final choice will be the model that maximizes equation (8) and that gives a ratio greater than 1.

**Discussion & Conclusion**

By drawing on the fundamental and necessary assumptions made by M. F. M. Osborne in order to describe the probability distribution of subjective change in prices, we tried to show that first considering the equiprobability of the opportunity of profit and loss between market participants, and by having a frequentist interpretation of the probability distribution of stock market changes, led to a certain misconception of the empirical reality, and could not fully explain the existing asymmetries within stock returns distribution leading to what we called the Osborne paradox.

We then propose two thought experiments illustrating by numerical computation in order to understand the reason of Osborne's assessments, and paradox.

By drawing on principles of the second law of Thermodynamics we understood that first,

Osborne's conclusion was valid and correct and second that the presence of weak asymmetries was due to an update in posterior probabilities of traders and investors. We also show that the stock market returns were respecting the second law of Thermodynamics with the increase in entropy until reaching an equilibrium state. This explaining the convergence of the long-term mean returns to zero as Osborne stipulated.

However, we also showed that although Osborne's conclusions remain true, it relies on a strong assumption that is not confirmed empirically namely that the stock market returns are a conservative system.

We demonstrated that the stock market returns must be considered as a dissipative system that exhibits spatio-temporally inhomogeneous dynamics, constituting of an ensemble of local macro-states (Coarse-grained) each in local equilibrium with different variance of posterior probability, in constant rotation and update due to new injection and subtraction of Coarse-grained with respective and different statistics metrics over time. And by superposing the ensemble of coarse-grained statistics, the stock market returns eventually exhibit fluctuating volatility, non-Gaussian distributions with fat tails, as Osborne stipulated.

We finally propose a theoretical approach to estimate the superposition of statistics of the ensemble of local marco-states via Bayesian Inference.

A future paper will propose a Bayesian hypothesis testing and model comparison in order to determine which superposition of statistics describes best the SP500 volatility and returns distributions using high-frequency data.